\documentclass[12pt]{article}
\usepackage{graphics}
\usepackage{amssymb}
\usepackage{amsmath}
\usepackage{upgreek}

\begin{document}

{\noindent \bf \Large Rip Brane Cosmology from 4d Inhomogeneous Dark Fluid Universe}

\bigskip

\begin{center}

I. Brevik\footnote{iver.h.brevik@ntnu.no}

\bigskip
Department of Energy and Process Engineering, Norwegian University of Science and Technology, N-7491 Trondheim, Norway

\bigskip
V. V. Obukhov, A. V. Timoshkin

\bigskip

Tomsk State Pedagogical University, 634061 Tomsk, Russia

\bigskip

Ye. Rabochaya

\bigskip
Eurasian National University, Astana, Kazakhstan

\bigskip
\today

\end{center}

\bigskip
\begin{abstract}

	Specific dark energy models with linear inhomogeneous time-dependent equation of state, within the framework   of 4d Friedman-Robertson-Walker (FRW) cosmology, are investigated. It is demonstrated that the choice of such 4d inhomogeneous fluid models may lead to a {\it  brane} FRW cosmology without any explicit account of higher dimensions at all. Effectively, we thus obtain a brane dark energy universe without introducing the brane concept explicitly. Several examples of brane Rip cosmology arising
from 4d inhomogeneous dark fluid models are given.
\end{abstract}

\section{ Introduction}
The discovery of the accelerating universe has led to the appearance of new ideas/solutions in cosmology \cite{riess98,perlmutter99}. This mysterious cosmic acceleration can be explained via the introduction of dark fluid (see Ref. [3] for recent review) or via modification of gravity itself (for review, see Ref.~\cite{nojiri06}). According to astronomical observations dark energy currently accounts for some 73\% of the total mass/energy of the universe and only 27\% of a combination of dark matter and baryonic matter. Dark energy proposed to explain the cosmic acceleration should have the strong negative pressure and/or negative entropy.

The equation of state dark energy parameter   is known to be negative:
\begin{equation}
w=\frac{p_D}{\rho_D}<0, \label{1}
\end{equation}
where  $\rho_D$  is the dark energy and  $p_D$  is the dark pressure.

According the present observational data value being  $w=-1.04_{-0.10}^{+0.09}$  \cite{nakamura10}. For a universe filled with phantom energy ($w<-1$   case) there are many possible new scenarios for the end of such universe. Phantom dark energy can lead to a Big Rip future singularity \cite{caldwell02,nojiri05,nojiri05A,nojiri03},  where the scale factor becomes infinite at a finite time in the future. Another possible scenario is a sudden (Type II) singularity \cite{shtanov02},  where the scale factor is finite at the Rip time(for general classification of finite-time future singularities see, Ref. \cite{nojiri05}). However, a final evolution without singularities is also possible. It occurs in modified gravity where higher-derivative term not only unifies inflation with dark energy  \cite{nojiri03A} but also removes the future singularity \cite{nojiri06}. On the other hand,
if the parameter $w$ asymptotically tends to $-1$, and the energy density increases with time or remains constant, no finite-time future singularity will be ever formed \cite{frampton11,frampton12,astashenok12,brevik11}. This is true also  if the universe starts to decelerate in the far future.

	If the cosmic energy density  remains constant or  increases monotonically  in the future, then all the possible fates of our universe can be divided into four categories based on the time asymptotic regimes of the Hubble parameter   \cite{frampton11,astashenok12}:

\bigskip

1.	Big Rip: $H(t)\rightarrow \infty$,   when  $t\rightarrow t_{\rm rip} <\infty$;

2.	Little Rip:  $H(t)\rightarrow \infty$  when  $t\rightarrow \infty$;

3.	Cosmological constant: $H(t) =$ const.;

4.	Pseudo-Rip:  $H(t)\rightarrow H_\infty <\infty$  when  $t\rightarrow \infty$,
with $H_\infty$   a constant.

\bigskip

	In this paper some examples of dark energy models of brane Rip cosmology from 4d FRW cosmology will be considered. Choosing appropriate values of the parameters   $w$ and  $\Lambda$ in the equation of state in standard FRW cosmology the Rip brane cosmology is obtained.

\section{Review of brane FRW cosmology}

We consider the simplest brane model in which space-time is homogeneous and isotropic along the three spatial dimensions, this being our 4-dimensional universe as an infinitesimally thin wall with constant spatial curvature, embedded in a 5-dimensional space-time \cite{sahni11,langlois03}.

The FRW metric on the brane is
\begin{equation}
ds^2|_{y=0}=-dt^2+a^2(t,0)\gamma_{ij}dx^idx^j, \label{2}
\end{equation}
where $\gamma_{ij}$ is the maximally symmetric 3-dimensional metric; $a$ is the scale factor.

The energy conservation equation is
\begin{equation}
\dot{\rho}_b+3\frac{\dot a}{a}(\rho_b+p_b)=0, \label{3}
\end{equation}
where $\rho_b$ and $p_b$ are the total brane energy density and pressure, respectively.

Now, let $\rho_b=\rho+\lambda$, where $\lambda$ is the brane tension. For the Hubble parameter we have the following equation
\begin{equation}
H^2=\frac{\rho}{3}\left( 1+\frac{\rho}{2\lambda}\right). \label{4}
\end{equation}
When $\rho \ll |\lambda|$, Eq.~(\ref{4}) differs insignificantly from the FRW equation. One can actually assume that in our epoch $\rho/2\lambda \ll 1$, and thus there is no significan difference between the brane model and conventional FRW cosmology.

The equation of state (EoS) for dark energy is taken in the form
\begin{equation}
p_D=-\rho_D-f(\rho_D), \label{5}
\end{equation}
where $f(\rho_D)$ is a function of the energy density. The case $f(\rho_D) >0$ corresponds to $w<-1$, whereas the case $f(\rho_D)<0$ corresponds to $w>-1$.

The EoS formalism for dark energy models on the brane was considered in Ref.~\cite{astashenok12B}. Assuming that $\rho_D \gg \rho_m$ where $\rho_m$ is the energy density of dark matter, one obtains the following link between time and energy density:
\begin{equation}
t(\rho_D)-t_0=\frac{1}{\sqrt 3}\int_{\rho_{D_0}}^{\rho_D} \frac{d\rho}{\sqrt {\rho}\left( 1+\frac{\rho}{2\lambda}\right)^{1/2}f(\rho)}, \label{6}
\end{equation}
where $t_0$ is the present time.

The scale factor as a function of the dark energy density obeys the same relation as in FRW cosmology:
\begin{equation}
a=a_0\exp \left(\frac{1}{3}\int_{\rho_{D_0}}^{\rho_D} \frac{d\rho}{g(\rho)}\right). \label{7}
\end{equation}

In the case of a {\it positive} tension, one has the following possibilities:

\bigskip

1. If the integral (\ref{6}) converges while (\ref{7}) diverges, there occurs a Big Rip.

2. If both integrals (\ref{6}) and (\ref{7}) diverge when $\rho_D \rightarrow \infty$, then a Little Rip occurs.

3. Asymptotic de Sitter expansion is realized if $g\rightarrow 0$ for $\rho_D \rightarrow \rho_{D_f}$, and the integral (\ref{6}) diverges.

4. The is a type III singularity if both integrals converge when $\rho_D\rightarrow \infty$.

5. If $g(\rho_D)\rightarrow \infty$ for $\rho_D\rightarrow \rho_{D_f}$, the universe ends its existence in a sudden future singularity.

\bigskip

If the tension is {\it negative}, the following ways of evolution are possible:

\bigskip
1. Asymptotic de Sitter expansion if $g(\rho_D)\rightarrow 0$ for $\rho_D \rightarrow \rho_{D_f}$.

2. Asymptotic breakdown (the rate of the universe tends to  zero), if $g(\rho_D)\rightarrow 0$ for $\rho_D\rightarrow 2\lambda$.

3. Sudden future singularity, if $g(\rho_D)\rightarrow \infty$ when $\rho_D\rightarrow \rho_{D_f}$.

\section{Examples of brane rip cosmology from 4d inhomogeneous dark fluid}

We consider now examples of dark energy models of brane Rip cosmology corresponding to the Little Rip case, the asymptotic de Sitter regime, and the Big Freeze singularity from 4d FRW cosmology. For simplicity it will be assumed that the universe consists of dark energy only.

\subsection{Little Rip case}

Let us consider a brane Little Rip model where the scale factor $a$ is given as \cite{astashenok12}
\begin{equation}
a(t)=a_0\exp \left[\frac{\lambda}{3\alpha^2}\cosh \left( \sqrt{\frac{3}{2\lambda}}\alpha^2t\right)\right]. \label{8}
\end{equation}
This corresponds to setting $f(\eta)=\alpha^2=$constant.
Here it is natural to associate $t=0$ with the present time, so that $a_0$ becomes the present-time scale factor.

The Friedman equation for a spatially flat universe is
\begin{equation}
\rho=\frac{3}{\kappa^2}H^2, \label{9}
\end{equation}
where $\rho$ is the energy density density, $H=\dot{a}/a$ the Hubble parameter, $a(t)$ the scale factor, and $\kappa^2=8\pi G$ with $G$ denoting  Newton's gravitational constant.

We assume that our universe is filled with an ideal fluid (dark energy) obeying an inhomogeneous equation of state (see Refs.~\cite{nojiri05,brevik04} for the general case):
\begin{equation}
p=w(t)\rho+\Lambda(t), \label{10}
\end{equation}
where $w(t)$ and $\Lambda(t)$ are time-dependent parameters and $p$ the pressure.

Let us write down the energy conservation law
\begin{equation}
\dot{\rho}+3H(\rho+p)=0. \label{11}
\end{equation}
The Hubble parameter is
\begin{equation}
H=\sqrt{\frac{\lambda}{6}}\sinh \left(\sqrt{\frac{3}{2\lambda}}\,\alpha^2t\right). \label{12}
\end{equation}
The derivative of $\rho$ with respect to cosmic time $t$ is
\begin{equation}
\dot{\rho}=3\alpha^2 \cosh \left(\sqrt{\frac{3}{2\lambda}}\,\alpha^2t\right) H. \label{13}
\end{equation}
Taking into account Eqs.~(\ref{10})-(\ref{13}) we obtain
\begin{equation}
\alpha^2\cosh \left(\sqrt{\frac{3}{2\lambda}}\, \alpha^2t\right) +\frac{\lambda}{2}[1+w(t)]\sinh^2\left( \sqrt{\frac{3}{2\lambda}}\,\alpha^2t\right) +\Lambda(t)=0. \label{14}
\end{equation}
Solving with respect to $\Lambda(t)$,
\begin{equation}
\Lambda(t)= -\left\{ \alpha^2\cosh \left(\sqrt{\frac{3}{2\lambda}}\, \alpha^2t\right) +\frac{\lambda}{2}[1+w(t)]\sinh^2\left( \sqrt{\frac{3}{2\lambda}}\,\alpha^2t\right)\right\},
\label{15}
\end{equation}
and choosing the parameter $w(t)$ as
\begin{equation}
w(t)=-1-\frac{2}{\lambda \sinh^2\left(\sqrt{\frac{3}{2\lambda}}\,\alpha^2t\right)}, \label{16}
\end{equation}
we find the cosmological "constant" to be
\begin{equation}
\Lambda(t)=1-\alpha^2\cosh \left( \sqrt{\frac{3}{2\lambda}}\,\alpha^2 t\right). \label{17}
\end{equation}
If $t\rightarrow \infty$, then the parameter $w$ tends asymptotically to $-1$ and the energy density increases monotonically with time. No finite-time future singularity is formed.

Consequently, if we assume an ideal fluid obeying an equation of state (\ref{10}) and (\ref{15}), then we obtain a solution realizing Little Rip on the brane, from the standpoint of 4d FRW cosmology.

\subsection{Asymptotic de Sitter regime}

We will now consider the situation where the brane has a {\it negative} tension ($\lambda <0$), corresponding to the universe expanding in a quasi-de Sitter regime.

The scale factor increases with time as \cite{astashenok12B}
\begin{equation}
a(t)=a_0\exp \left\{ \frac{\beta^2}{2}\left[ \sqrt{1+\left(\tan \eta_0+\frac{\alpha t}{\beta}\right)^2}-\frac{1}{\cos \eta_0}\right]\right\}. \label{18}
\end{equation}
In this equation the dimensionless parameter $\beta^2=2|\lambda|/(3\alpha^2)$ has been introduced, and also $\eta_0=\sqrt{3/2\lambda}\,\alpha^2t_0$, where $t_0$ is the present time.

The Hubble parameter becomes
\begin{equation}
H=\frac{\alpha \beta}{2}\frac{\tan \eta_0+\frac{\alpha}{\beta}t}{\sqrt{ 1+\left( \tan \eta_0+\frac{\alpha }{\beta}t\right)^2}}. \label{19}
\end{equation}
When $t\rightarrow \infty$, $H\rightarrow \alpha \beta/2$.  Thus the expression (\ref{19}) asymptotically tends to the de Sitter solution.

We take the derivative of the energy density with respect to cosmic time,
\begin{equation}
\dot{\rho}=\frac{3\alpha^2 H}{\sqrt{ 1+\left(\tan \eta_0+\frac{\alpha}{\beta}t\right)^2}}, \label{20}
\end{equation}
and from Eqs.~(\ref{10}), (\ref{11}), (\ref{19}) and (\ref{20}) we then obtain the energy conservation law,
\begin{equation}
\frac{2\alpha}{\beta}\frac{H}{\tan \eta_0+\frac{\alpha}{\beta}t}+3(1+w)H^2+\Lambda=0, \label{21}
\end{equation}
where $t \neq -(\beta/\alpha)\tan \eta_0$.

Let us solve Eq.~(\ref{21}) with respect to $\Lambda(t)$,
\begin{equation}
\Lambda(t)=-H\left[ 3H(1+w)+\frac{2\frac{\alpha}{\beta}}{\tan \eta_0+\frac{\alpha}{\beta}t}\right]. \label{22}
\end{equation}
Now writing the parameter $w(t)$ in the form
\begin{equation}
w(t)=-1-\frac{\delta}{3H}, \label{23}
\end{equation}
with $\delta$ a positive constant, we obtain from Eq.~(\ref{22})
\begin{equation}
\Lambda(t)=-H\left[ \frac{2\frac{\alpha}{\beta}}{\tan \eta_0+
\frac{\alpha}{\beta}t}-\frac{\delta}{3H}\right]. \label{24}
\end{equation}
If $t\rightarrow +\infty$, then $\Lambda \rightarrow \delta/3$.

Thus we have presented the appearance of the asymptotic de Sitter regime on the brane, from 4d FRW cosmology.

\subsection{ Big Freeze singularity cosmology}

Let us assume that $\lambda >0$. There are then two Big Freeze singularities: one in the past ($t\rightarrow -\infty$) and another in the future ($t\rightarrow +\infty$). The universe begins its existence at $t_{\rm in}=-\frac{1}{\alpha^2}\sqrt{ \frac{2\lambda}{3}}$ and ends at $t_f= \frac{1}{\alpha^2}\sqrt{ \frac{2\lambda}{3}}$. The scale factor is
\begin{equation}
a(t)=a_f\exp \left[ -\frac{\lambda}{3\alpha^2}\left( 1-\frac{3\alpha^4t^2}{2\lambda}\right)^{1/2}\right], \label{25}
\end{equation}
where $a_f$ is the final scale factor. This expression shows that the universe will contract in the time interval $t_{\rm in} <t<0$ and will thereafter expand.

The Hubble parameter is
\begin{equation}
H=\frac{\alpha^2t}{2\sqrt{1-\frac{3\alpha^4t^2}{2\lambda}}}. \label{26}
\end{equation}
We now find
\begin{equation}
\dot{\rho}=3H\frac{\alpha^2}{\left( 1-\frac{3\alpha^4t^2}{2\lambda}\right)^{3/2}}. \label{27}
\end{equation}
Using Eqs.~(\ref{10}), (\ref{11}), (\ref{26}) and (\ref{27}) we can rewrite the energy conservation equation as
\begin{equation}
\frac{\alpha^2}{\left( 1-\frac{3\alpha^4t^2}{2\lambda}\right)^{3/2}}+\frac{3}{4}\alpha^2(1+w)\frac{t^2}{1-\frac{3\alpha^4t^2}{2\lambda}}+\Lambda=0. \label{28}
\end{equation}
We may solve for $\Lambda(t)$,
\begin{equation}
\Lambda(t)=-\frac{\alpha^2}{1-\frac{3\alpha^4t^2}{2\lambda}}\left[ \frac{3}{4}\alpha^2(1+w)t^2+\frac{1}{\sqrt{ 1-\frac{3\alpha^4t^2}{2\lambda}}}\right]. \label{29}
\end{equation}
If we assume that the parameter $w(t)$ has the following time dependence,
\begin{equation}
w(t)=-1-\frac{1}{\sqrt{1-\frac{3}{2}\frac{\alpha^4}{\lambda}t^2}}, \label{30}
\end{equation}
we obtain
\begin{equation}
\Lambda(t)=\frac{\alpha^2\left( 1+\frac{3}{4}\alpha^2t^2\right)}{\left( 1-\frac{3\alpha^4t^2}{2\lambda}\right)^{3/2}}. \label{31}
\end{equation}
Thus, we have explored the appearance of a Big Freeze (Type III) singularity on the brane, from 4d FRW cosmology, where brane effects are mimicked by an inhomogeneous fluid.

\section{Conclusion}

Several brane dark energy models have been analyzed in accordance with the usual 4d FRW cosmology, including asymptotic de Sitter evolution, Little Rip behavior, and Big Freeze singularity. The brane contribution has been effectively taken into account via the corresponding choice of an inhomogeneous dark fluid. In other words, we have obtained a brane dark energy universe without introducing the brane concept explicitly.

A prediction of the future evolution of the universe is closely associated with the chosen EoS for the dark energy component. Choosing parameters $w(t)$ and $\Lambda(t)$ in the inhomogeneous time-dependent EoS, the effective FRW cosmology on the brane is obtained.

\vspace{1.5cm}
{\bf Acknowledgments}

\bigskip

\noindent This work has been supported by project 2.1839.2011 of Min. of Education and Science (Russia) and LRSS project 224.2012.2 (Russia). We are grateful to Professor Sergei Odintsov for clarifying discussions.

\end{document}